\documentclass[11pt]{article}

\usepackage[utf8]{inputenc}
\usepackage[T1]{fontenc}
\usepackage{lmodern}
\usepackage[margin=1in]{geometry}
\usepackage{microtype}
\usepackage{graphicx}
\usepackage{booktabs}
\usepackage{tabularx}
\usepackage{array}
\usepackage{amsmath}
\usepackage{enumitem}
\usepackage{titlesec}
\usepackage{abstract}
\usepackage[numbers,sort&compress]{natbib}
\usepackage[colorlinks=true,linkcolor=black,citecolor=black,urlcolor=blue]{hyperref}
\usepackage{orcidlink}

\newcolumntype{L}[1]{>{\raggedright\arraybackslash}p{#1}}
\titleformat{\section}{\normalfont\large\bfseries}{\thesection}{0.6em}{}
\titleformat{\subsection}{\normalfont\normalsize\bfseries\itshape}{\thesubsection}{0.6em}{}

\title{\bfseries The Deployment Wall: A Diagnostic Framework and Instrument for Enterprise AI in the Deployment Era}

\author{
  Fabricio F. Costa\thanks{Correspondence: \texttt{fcosta@aix4all.com}. The views expressed are the author's own.}\\
  \small HCLTech\\
  \small \texttt{fcosta@aix4all.com}
}
\date{\today}

\begin{document}
\maketitle

\begin{abstract}
\noindent Enterprise investment in generative artificial intelligence (AI) tripled in a single year to roughly US\$37\,billion, yet independent field research finds that about 95\% of enterprise generative-AI pilots deliver no measurable profit-and-loss impact. We argue that the dominant explanation---that models are not yet capable enough---is mistaken, and that enterprise AI has entered a \emph{Deployment Era} in which advantage derives not from model intelligence but from the removal of the organizational and architectural friction that prevents a capable model from reaching production. Building on the software-engineering literature on technical debt and machine-learning deployment, and on a structured synthesis of independent field studies, we make the diagnosis operational. We introduce three linked constructs and one measurement instrument: the \emph{Deployment Wall}, a six-stage value-leak model that mechanically reproduces observed survival rates; the \emph{Seam Index}, a reproducible 0--12 diagnostic that scores any platform by how many of six recurring friction ``seams'' it removes natively rather than leaving to the adopter; and \emph{Deployment Debt}, a construct that reframes unresolved friction as a compounding, quantifiable liability. We specify a scoring protocol with evidence anchors so the instrument can be applied consistently, illustrate it on a worked platform-selection example, and derive six falsifiable propositions with a research agenda for validation. The framework converts an eight-figure platform decision from a benchmark comparison into an architecture comparison.
\end{abstract}


\section{Introduction: The End of the Intelligence Era}
\begin{figure}[t]
\centering
\includegraphics[width=0.82\linewidth]{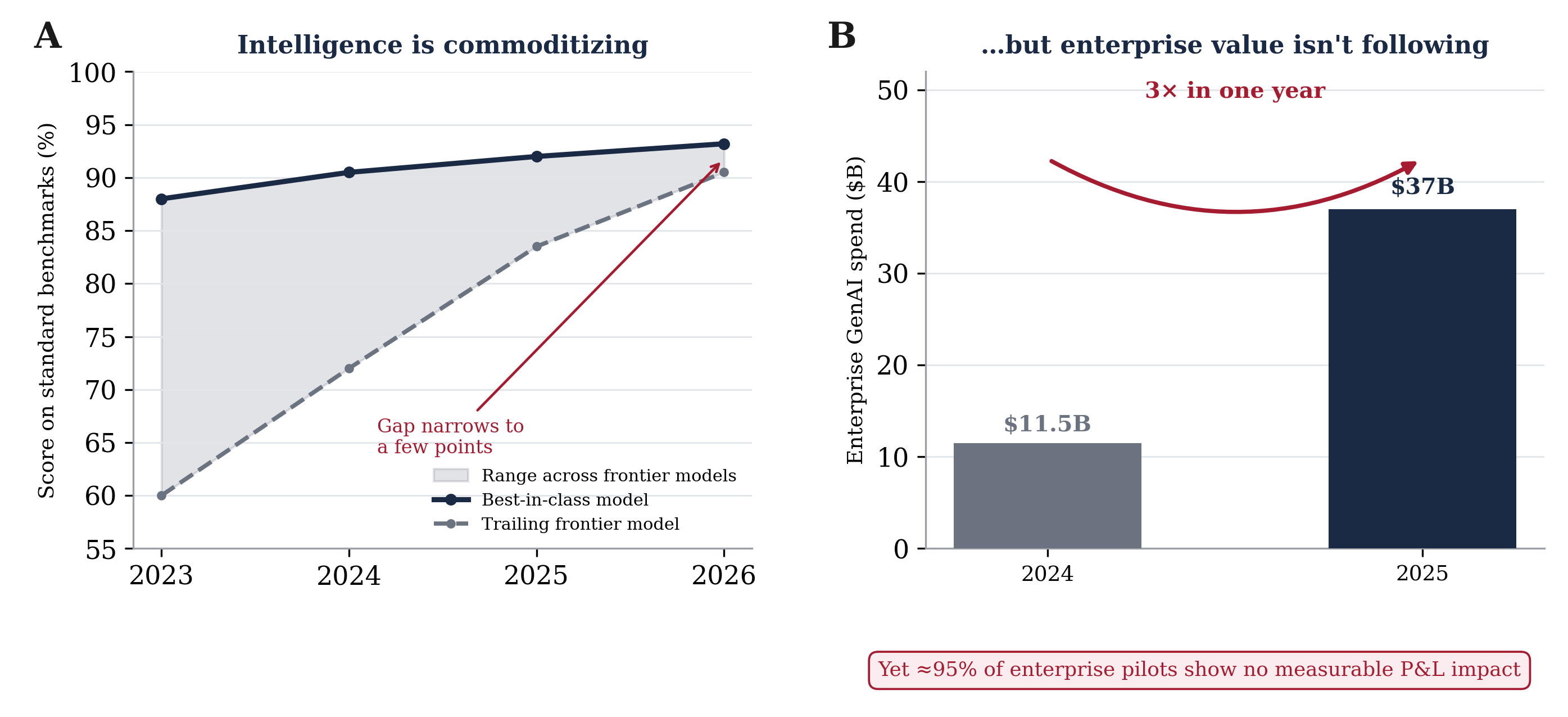}
\caption{Intelligence is commoditizing, but enterprise value is not following. (A) Frontier-model benchmark scores converge as the range across leading models narrows (illustrative, based on public evaluations). (B) Enterprise generative-AI spending tripled in one year even as $\approx$95\% of pilots showed no measurable P\&L impact.}
\label{fig:converge}
\end{figure}

For three years, the organizing logic of enterprise artificial intelligence (AI) has been a race for intelligence. Vendors competed on parameter counts and benchmark scores; buyers tracked leaderboards; boards asked whether their organization had access to the most capable model. That race is effectively over. Frontier models now converge at the top of every standard benchmark, and the gap between the best and the fifth-best model has narrowed to fractions of a point on most public evaluations (Fig.~\ref{fig:converge}). Intelligence, in other words, is commoditizing: the marginal capability that once separated market leaders from the field is increasingly available to any organization willing to pay per token.

This convergence is not a claim that models have stopped improving; they continue to advance. It is a claim about marginal competitive value. When the gap between the leading model and the fifth-ranked model is a fraction of a point on evaluations that few enterprise tasks resemble, the model chosen is unlikely to be the variable that decides whether an initiative succeeds. The decisive variables lie elsewhere---in the organization that must absorb the model---which is exactly where the following sections direct attention.

Against this backdrop, the enterprise results are difficult to reconcile with the level of investment. Enterprise spending on generative AI tripled in a single year to roughly US\$37\,billion \citep{menlo2025}, and yet independent field research from MIT's Project NANDA reports that approximately 95\% of enterprise generative-AI pilots produce no measurable profit-and-loss impact \citep{nanda2025}. The reflexive interpretation---that models are not yet good enough, and that the next, smarter model will unlock returns---is, we argue, precisely the wrong diagnosis. The same research is explicit that the divide between the few organizations capturing value and the many that are not is determined by approach, not by model quality \citep{nanda2025}. The barrier is not intelligence. It is deployment.

This paper develops that claim into a managerial framework and, crucially, into a \emph{measurement instrument}. Our contribution is deliberately design-oriented rather than argumentative: we do not merely assert that deployment is the bottleneck, we make the bottleneck locatable, measurable, and costable. Specifically, the paper contributes (i) the \emph{Deployment Wall}, a six-stage value-leak model that reproduces, mechanically, the survival rate that field studies observe; (ii) the \emph{Seam Index}, a 0--12 diagnostic instrument, accompanied by a reproducible scoring protocol with evidence anchors, that converts a platform decision from a benchmark comparison into an architecture comparison; (iii) \emph{Deployment Debt}, a construct that reframes unresolved friction as a compounding financial, organizational, and competitive liability suitable for board-level discussion; and (iv) six falsifiable propositions and a research agenda through which the framework can be empirically validated. Building on these, we argue that the durable competitive moat has migrated from the model layer to the seam layer, and that this moat is distributed across hyperscalers that each own different seams.

The stakes of getting this diagnosis right are large and asymmetric. If the binding constraint is intelligence, then patience is the correct strategy and today's investment is merely early. If, as we argue, the binding constraint is deployment, then patience is the most expensive option available, because the organizations that build deployment capability now compound an advantage that late movers cannot easily buy. Misdiagnosis is therefore not a neutral error; it systematically favors inaction at precisely the moment when action is cheapest.

The argument is deliberately platform-neutral and decision-oriented. It is written for the senior technology and engineering manager who must allocate capital and organizational attention under uncertainty, and who needs a durable mental model---and a usable instrument---rather than a product recommendation. The remainder of the paper is organized as follows. Section~\ref{sec:related} situates the work in prior research on technology adoption, machine-learning deployment, and enterprise-AI value. Section~\ref{sec:method} describes our design method and evidentiary basis. Sections~\ref{sec:wall}--\ref{sec:moat} develop the constructs, the instrument, and the moat argument. Section~\ref{sec:practice} translates the framework into managerial action, and Sections~\ref{sec:props}--\ref{sec:concl} present propositions, limitations, and conclusions.

\section{Related Work}
\label{sec:related}
The observation that technological capability and organizational value are distinct is among the most durable findings in the management of technology. We position our contribution against four literatures and identify the gap it addresses.

\subsection{Technology adoption is an organizational process}
Rogers' diffusion-of-innovations theory established that adoption proceeds through stages of knowledge, persuasion, decision, implementation, and confirmation, and that most failures occur not at the decision to adopt but during implementation and confirmation \citep{rogers2003}. The technology--organization--environment (TOE) framework holds that adoption outcomes are jointly determined by technological, organizational, and environmental factors, so that a superior technology embedded in an unprepared organization underperforms an inferior technology embedded in a ready one \citep{tornatzky1990}. Cohen and Levinthal's absorptive capacity---an organization's ability to recognize, assimilate, and apply new external knowledge---predicts that firms differ systematically in their capacity to convert the same external technology into performance \citep{cohen1990}. Because the external technology (the frontier model) is now broadly available, variance in outcomes must be explained by variance in absorptive capacity; our seam analysis (Section~\ref{sec:seams}) can be read as an operational decomposition of that capacity for AI.

\subsection{Technical debt and the software engineering of ML deployment}
Our core mechanism is adapted from software engineering. Technical debt describes the future cost incurred when an expedient near-term implementation must later be reworked; unpaid, it compounds and eventually dominates maintenance cost \citep{cunningham1992}. Sculley et al.\ extended this to machine learning, showing that the model code is a small fraction of a real ML system, which is dominated by configuration, data collection and verification, serving infrastructure, and monitoring, and that ML systems incur \emph{additional}, hidden forms of debt through entanglement and undeclared consumers \citep{sculley2015}. Subsequent work documents the same pattern empirically: deploying ML is dominated by data, integration, and operational challenges rather than modeling \citep{paleyes2022,lwakatare2020}; the data stage in particular is chronically undervalued \citep{sambasivan2021}; and the discipline of MLOps has emerged precisely to manage this operational surface \citep{kreuzberger2023,amershi2019,serban2020}. Notably, this literature has produced \emph{rubric-style} instruments---most directly the ML Test Score, a checklist for production readiness and technical-debt reduction \citep{breck2017}. Our Seam Index is in this tradition, but it operates one level up: rather than scoring an individual system's engineering hygiene, it scores a \emph{platform choice} by the organizational friction it removes, making it a decision instrument for buyers rather than a readiness checklist for builders.

\subsection{Enterprise AI value, capability, and governance}
A parallel information-systems literature studies why AI investment does or does not convert into value. Recent syntheses catalog the organizational determinants of AI value \citep{enholm2022,benbya2020}, treat AI capability as a firm-level construct that must be built rather than bought \citep{mikalef2021}, and frame the governance of AI as a distinct managerial problem \citep{berente2021}. Practitioner-facing scholarship reaches a consistent conclusion: value comes from redesigning work and organizations around AI, not from the technology alone \citep{davenport2018,fountaine2019,wilson2018}. Opinion and position papers on generative AI specifically identify governance, data, and skills as central concerns \citep{dwivedi2023}, and engineering-management work enumerates the applications, challenges, and adoption requirements of generative AI in organizations \citep{alkfairy2025}. We contribute to this literature a compact, measurable instrument that operationalizes ``organizational readiness'' as six scorable seams.

\subsection{AI strategy, complementary assets, and competitive advantage}
Finally, the strategy literature explains where advantage accrues once a general-purpose technology diffuses. Teece's theory of complementary assets predicts that when a core technology is imitable, returns flow to whoever controls the co-specialized assets required to commercialize it \citep{teece1986,teece2007}. Iansiti and Lakhani argue that competitive advantage in AI comes from the operating model---the ``AI factory''---rather than from any single model \citep{iansiti2020}, and the economics-of-AI literature frames adoption as a question of complements to cheap prediction \citep{agrawal2018}. Our moat argument (Section~\ref{sec:moat}) is a specialization of this logic: when models commoditize, the co-specialized assets are the seams, and they are distributed across providers.

\subsection{The gap}
Prior work establishes that implementation, absorptive capacity, ML operations, and organizational readiness matter, and it has produced readiness rubrics for individual ML systems. What the literature has not provided is a compact, reproducible, decision-grade instrument that a manager can apply \emph{before committing capital} to compare candidate platforms by the friction they remove from a specific enterprise environment. That is the gap this paper addresses.

\section{Method and Evidentiary Basis}
\label{sec:method}
This paper is a conceptual and design-oriented contribution rather than an empirical study, and it is presented as such. Its frameworks and instrument were derived and are supported by two complementary sources of evidence, and the boundaries of each are stated explicitly so that readers can weigh the claims appropriately.

\textbf{Field observation.} The constructs were developed inductively from the author's direct professional experience deploying frontier models into production across organizations of varying size---first as an independent consultant serving enterprises, mid-market firms, and startups, and subsequently leading enterprise deployment engagements at a global technology services firm. The recurring pattern across these engagements---that initiatives stalled at a consistent set of organizational seams rather than at model capability---motivated the framework. This is participant-observation evidence: it is rich and longitudinal but non-random and not systematically sampled, and it is offered as the origin of the hypotheses, not as their statistical confirmation.

\textbf{Structured synthesis of independent studies.} To triangulate the field observations, the paper draws on a structured synthesis of publicly available studies from research organizations and market analysts, summarized in Table~\ref{tab:studies}, cross-checked against public company disclosures. These include large-sample surveys and field studies of AI outcomes \citep{nanda2025,rand2024,sandp2025,mckinsey2025,bcg2025}, enterprise-spend and market-share data \citep{menlo2025,ramp2026}, and seam-specific evidence on data, security, and governance readiness \citep{deloitte2025,layerx2025,ibm2024,k2view2025}. Where a figure derives from a single source it is attributed to that source; where several sources converge, that convergence is noted.

\begin{table}[t]
\centering
\small
\caption{Independent studies synthesized (outcomes and enterprise adoption).}
\label{tab:studies}
\begin{tabularx}{\linewidth}{L{3.3cm} L{3.4cm} X}
\toprule
\textbf{Source} & \textbf{Scope / sample} & \textbf{Headline finding} \\
\midrule
MIT Project NANDA (2025) \citep{nanda2025} & 150 interviews; $\sim$350 surveys; 300 deployments & $\sim$95\% of enterprise GenAI pilots show no measurable P\&L impact. \\
RAND (2024) \citep{rand2024} & 65 expert interviews & Most AI projects fail to reach production; causes are largely organizational. \\
S\&P Global (2025) \citep{sandp2025} & Survey of $\sim$1,000+ firms & Share of firms abandoning AI initiatives rose year over year. \\
McKinsey (2025) \citep{mckinsey2025} & Global survey ($\sim$1,900 respondents) & A minority redesign workflows; adoption, not model, gates value. \\
BCG (2025) \citep{bcg2025} & $\sim$1,000 executives & Only a minority capture significant value from AI at scale. \\
Menlo Ventures (2025) \citep{menlo2025} & Enterprise buyer survey & Spend tripled to $\sim$US\$37B; integration and maintenance dominate cost. \\
\bottomrule
\end{tabularx}
\end{table}

\textbf{Reproducibility.} The three named constructs (Deployment Wall, Seam Index, Deployment Debt) are original conceptual models; the Seam Index is accompanied by an explicit scoring protocol (Section~\ref{sec:index}) so that independent evaluators can apply it consistently. Several figures are explicitly labeled as illustrative models that encode the direction and rough magnitude of the cited evidence rather than a measured dataset; they communicate structure, not new measurements. The propositions in Section~\ref{sec:props} are stated in falsifiable form precisely so that subsequent empirical work can test them.

Two boundary conditions on generalizability should be stated. The framework is developed from, and most directly applicable to, medium-to-large organizations operating in regulated or data-sensitive environments, where the six seams are most binding; very small organizations with simple data estates and few compliance obligations may face materially lower seam burden, and for them model quality may indeed be the more decisive variable. The framework is also calibrated to the current generation of enterprise deployment; as tooling matures and platforms absorb more seams by default, the absolute height of the wall should fall even as its structure persists.

\section{The Deployment Wall: A Six-Stage Value-Leak Model}
\label{sec:wall}
Enterprise AI initiatives rarely fail in a single, visible collapse. They leak. Value drains away gradually as an initiative attempts to pass through a sequence of organizational gates, each of which removes a fraction of the initiatives that reach it. We model this as a wall built of six courses: (1)~model selection, (2)~integration, (3)~governance, (4)~workflow redesign, (5)~enterprise adoption, and (6)~realized business value. An initiative must clear every course; failure at any one arrests it.

The six courses are cumulative and ordered. \emph{Model selection} chooses a capable model and proves feasibility, usually the fastest and most visible stage. \emph{Integration} connects that model to enterprise systems of record and data sources, exposing the first hard seams. \emph{Governance} establishes the monitoring, auditability, and accountability required before any regulated organization will permit production use. \emph{Workflow redesign} rebuilds the surrounding process so that the capability changes how work is done rather than sitting beside it. \emph{Enterprise adoption} drives real usage across the intended population, which depends on trust, training, and incentives as much as on the interface. Finally, \emph{realized business value} requires that all of the above translate into a measurable operational or financial result. Each stage has a distinct owner, a distinct failure mode, and a distinct cost, and an initiative that clears five stages but stalls at the sixth has still delivered nothing to the profit-and-loss statement.

\begin{figure}[t]
\centering
\includegraphics[width=0.72\linewidth]{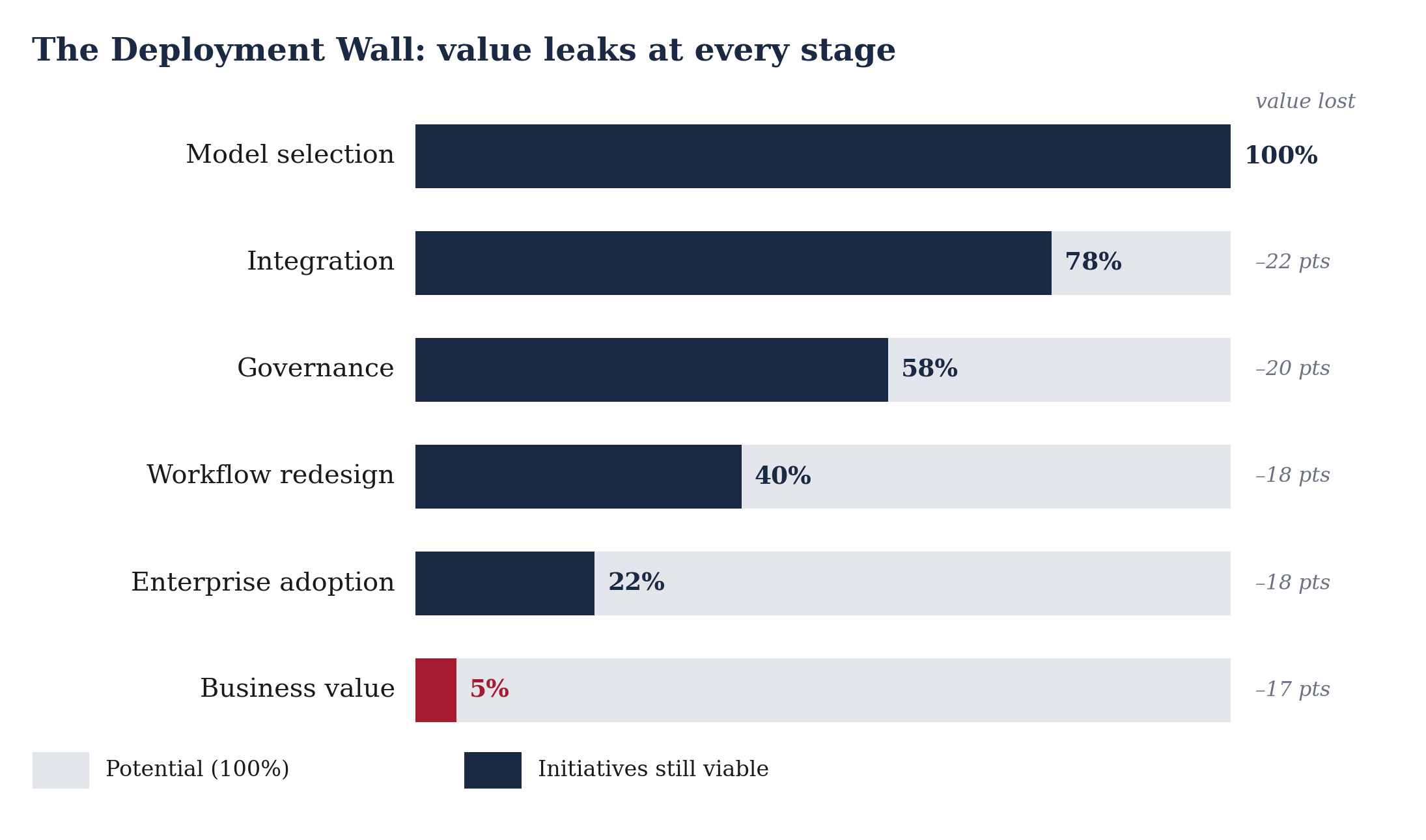}
\caption{The Deployment Wall. Cumulative share of initiatives surviving each successive stage (illustrative structure consistent with field survival rates).}
\label{fig:wall}
\end{figure}

Figure~\ref{fig:wall} renders the wall as a survival funnel. If one assumes even moderate attrition at each stage---reflecting the field evidence that most initiatives never traverse integration, governance, and workflow redesign---the cumulative survival rate at the final stage falls to the single digits, mechanically reproducing the small ``survivors' club'' that field research observes \citep{nanda2025}. The precise per-stage rates are illustrative; the structural point is not. When six independent gates each remove a portion of the pipeline, only a small fraction of initiatives can emerge with realized value, regardless of how capable the underlying model is.

\begin{figure}[t]
\centering
\includegraphics[width=0.78\linewidth]{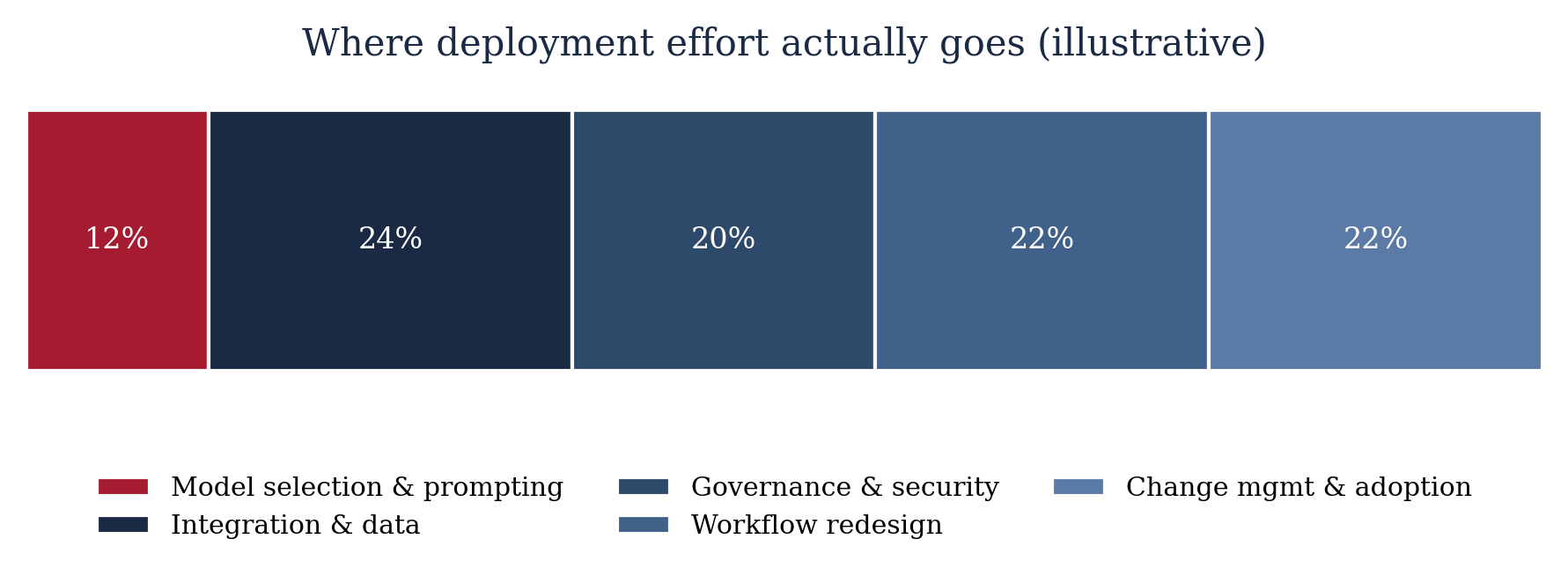}
\caption{Where deployment effort goes. Model selection is a small share of the total work that determines outcomes (illustrative).}
\label{fig:effort}
\end{figure}

The most consequential implication is visible in Figure~\ref{fig:effort}, which allocates the total effort of a representative deployment across activities. Model selection---the activity that consumes the greatest share of executive attention and vendor comparison---accounts for a small share of the work that determines success. The overwhelming majority of effort, and of failure risk, lies in the five courses that follow: connecting the model to enterprise data and systems, governing it, redesigning the work around it, and driving adoption. The model is table stakes; the wall is the work.

\section{The Six Seams Where Value Leaks}
\label{sec:seams}
If the wall describes where initiatives stall, the seams describe why. A seam is a boundary an AI capability must cross to function in an enterprise, and each seam is a place where friction accumulates. Six seams recur across engagements and are corroborated by independent evidence, summarized in Table~\ref{tab:seams}.

\begin{table}[t]
\centering
\small
\caption{The six seams where enterprise AI value leaks.}
\label{tab:seams}
\begin{tabularx}{\linewidth}{L{2.4cm} X X}
\toprule
\textbf{Seam} & \textbf{What must be crossed} & \textbf{Representative evidence} \\
\midrule
Data & Fragmented, poorly governed, or siloed enterprise data must be made accessible and trustworthy to the model. & Data quality and readiness cited as a leading barrier; large shares of effort consumed by data integration \citep{k2view2025,sambasivan2021}. \\
Identity \& access & The capability must respect role-based access so that it sees only what each user may see. & A majority of corporate generative-AI connections lack single sign-on, exposing an identity gap \citep{layerx2025}. \\
Security \& compliance & Outputs and data flows must satisfy security policy and sector regulation. & Security, privacy, and regulatory concern repeatedly rank among the top barriers to scaling \citep{deloitte2025}. \\
Governance & Monitoring, auditability, and accountability must exist before production use. & Only a minority of organizations report robust, enterprise-wide AI governance \citep{ibm2024,berente2021}. \\
Change management & People and processes must be redesigned, and skills built, for the new capability. & A minority of firms redesign workflows; adoption, not technology, gates value \citep{mckinsey2025,fountaine2019}. \\
Cost & Total cost of ownership---integration, maintenance, and inference---must be controlled over time. & Ongoing maintenance and integration costs dominate lifetime spend and erode return \citep{menlo2025,sculley2015}. \\
\bottomrule
\end{tabularx}
\end{table}

Each seam merits brief elaboration, because the character of the friction differs by seam. The \textbf{data seam} is where most initiatives first bleed time. A frontier model is only as useful as its access to trustworthy, well-governed enterprise data, yet in most organizations that data is fragmented across systems of record, data lakes, and departmental silos, with inconsistent quality and unclear ownership. Retrieval-augmented architectures shift this problem rather than remove it: they demand clean, permissioned, well-indexed sources, and assembling those sources routinely consumes the single largest share of a deployment's effort---a pattern the ML literature calls data cascades \citep{sambasivan2021,k2view2025}.

The \textbf{identity and access seam} determines whether a capable assistant is safe to expose to real users. In a compliant enterprise, what a system may retrieve or generate depends on who is asking; a model that ignores role-based access control is not deployable regardless of its intelligence. Yet a majority of corporate generative-AI connections have been found to lack single sign-on, meaning the identity layer that governs every other enterprise system is effectively absent for AI \citep{layerx2025}. Closing this seam requires wiring the model into existing identity providers and entitlement models---unglamorous work that nonetheless gates production.

The \textbf{security and compliance seam} governs whether outputs and data flows satisfy security policy and sector regulation. Prompt-injection exposure, data-egress risk, and sector rules on privacy and record-keeping each impose controls that must exist before, not after, launch; in regulated sectors, the penalty for getting this wrong is measured in enforcement actions, not embarrassment. Security and compliance concerns rank persistently among the top barriers organizations cite to scaling AI \citep{deloitte2025}.

The \textbf{governance seam} is the difference between a demonstration and a system a board will stand behind. It requires monitoring for drift and error, logging for auditability, human-escalation paths, and clear accountability for outcomes---capabilities that only a minority of organizations report having established enterprise-wide \citep{ibm2024}, and that the IS literature treats as a first-class managerial problem \citep{berente2021}. Without them, an initiative can function technically and still be unfit for production, because no one can answer for what it does.

The \textbf{change-management seam} is the seam most correlated with realized value and the one most often neglected. Technology deployed without redesigning the surrounding work, and without building skills and incentives for its use, produces novelty rather than impact; field research finds that only a minority of firms redesign workflows, and that adoption---not model capability---is what ultimately gates value \citep{mckinsey2025}. This seam is organizational, not technical, which is precisely why engineering-led initiatives so frequently underestimate it \citep{fountaine2019}.

The \textbf{cost seam} governs whether an initiative that works is also worth operating. Total cost of ownership is dominated not by the initial build but by ongoing integration, maintenance, monitoring, and inference spend, which accumulate for the life of the system \citep{menlo2025,sculley2015}. A capability that cannot be operated within a defensible cost envelope will be withdrawn regardless of its quality, and the organizations that ignore this seam early are the ones most surprised by it later.

A single engagement illustrates how decisively the seams, rather than the model, determine outcomes. A global bank built a strikingly capable regulatory-compliance assistant over a single weekend; the model was never in doubt. The initiative then spent nine months and roughly US\$2\,million attempting to map the assistant onto role-based access controls and legacy, siloed data before the rollout was abandoned. The capability existed on day two; the seams defeated it over the following three quarters. This is the Deployment Wall in miniature: capability at stage one, value never realized because integration, identity, and governance were not crossed. The engagement is instructive precisely because none of its difficulty was exotic. The bank did not lack talent, budget, or a capable model; it lacked crossed seams---and the seams it faced are the ordinary furniture of a regulated enterprise.

\section{External Validation: Why the Labs Are Hiring Armies of Consultants}
\label{sec:tell}
If a smarter model were the binding constraint, the organizations that possess the smartest models would simply ship them and capture the value directly. Their revealed behavior is the opposite. The frontier laboratories are racing to acquire deployment capacity---people and partnerships whose function is to cross the seams inside client organizations---at extraordinary scale (Table~\ref{tab:landgrab}). This is the clearest external validation of the thesis: the firms with the most to gain from a pure-intelligence explanation are themselves investing as though deployment, not intelligence, is the bottleneck.

\begin{table}[t]
\centering
\small
\caption{The consultant land-grab: frontier labs acquiring deployment capacity.}
\label{tab:landgrab}
\begin{tabularx}{\linewidth}{L{2.0cm} X L{3.3cm}}
\toprule
\textbf{Provider} & \textbf{Move} & \textbf{Signal} \\
\midrule
Anthropic & Alliance to deploy its models to roughly 470,000 seats, bundled with $\sim$15,000 certified consultants \citep{anthropic2025}. & Deployment capacity, not model access, is the scarce good. \\
OpenAI & Enterprise alliance program plus a multibillion-dollar dedicated deployment company \citep{openai2025}. & Investing directly in crossing client seams. \\
Google & Hundreds of thousands of trained consultants and a US\$750M partner fund for deployment \citep{google2025}. & Subsidizing the removal of friction at scale. \\
Microsoft & Forward-deployed engineering with global systems integrators \citep{microsoft2025}. & Distribution and integration treated as the moat. \\
\bottomrule
\end{tabularx}
\end{table}

Two implications follow. First, the market for AI consulting is migrating from implementing models---a comparatively small task---to redesigning organizations, which is where the seams live. Second, for the buyer, heavy consulting requirements are better read as a friction signal than as a mark of sophistication: if a platform requires a small army to deploy, its seam burden is high, and that burden is a cost the buyer will carry long after the engagement ends. As models commoditize, the differentiated and defensible work is not prompt engineering or model selection but the sustained, organization-specific labor of crossing seams. That is why the frontier labs are not merely licensing models to integrators but acquiring or building deployment organizations outright---they are moving to capture the part of the value chain where the scarce capability now lives.

\section{Deployment Debt: A Measurable, Compounding Liability}
\label{sec:debt}
For a finance audience, friction is easy to underweight because it presents as delay rather than as loss. Reframing it as debt---in direct analogy to technical debt in software \citep{cunningham1992,sculley2015}---corrects that error. \emph{Deployment Debt} is the accumulated, compounding cost of seams left unresolved, and it takes three forms.

\begin{figure}[t]
\centering
\includegraphics[width=0.74\linewidth]{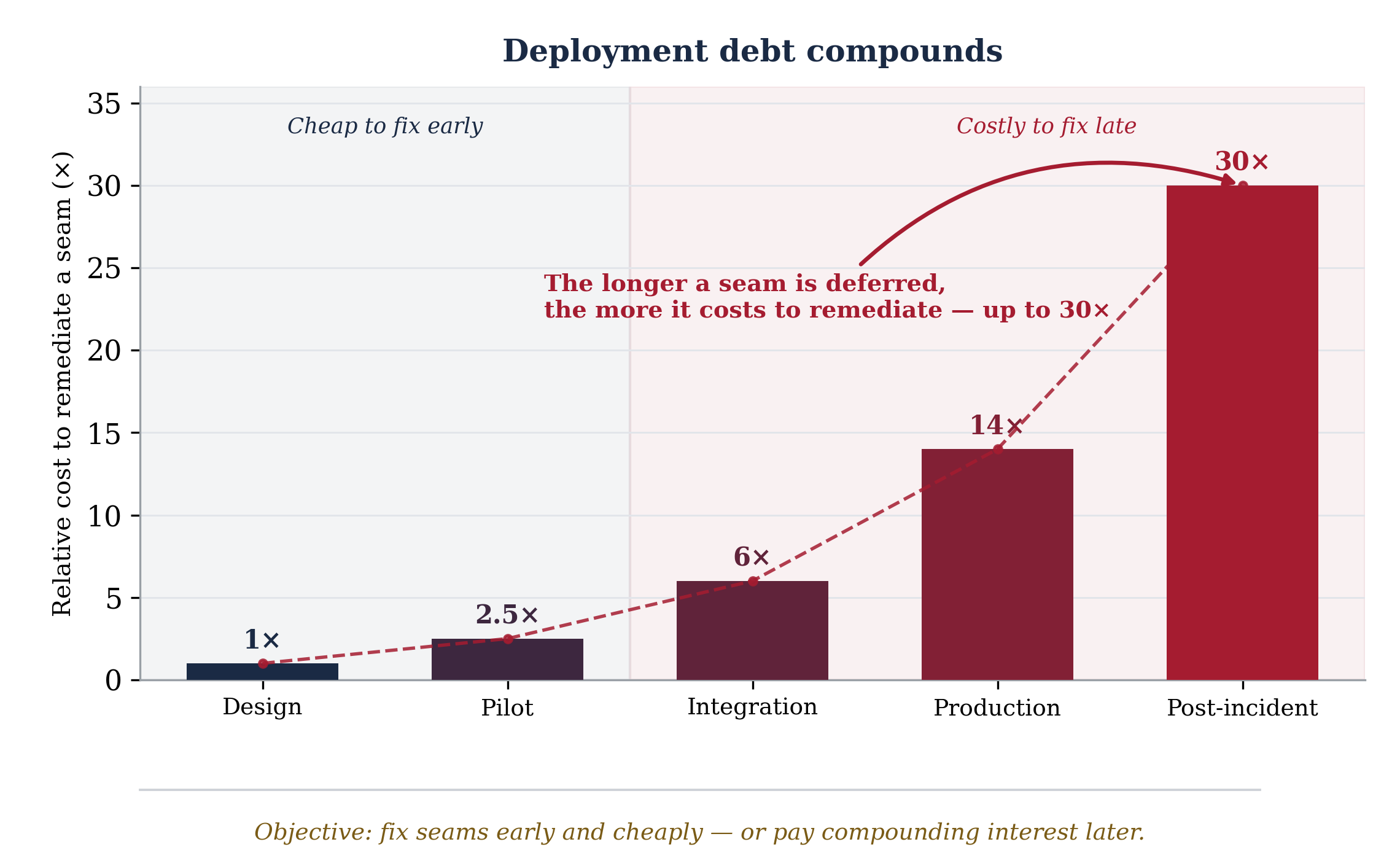}
\caption{Deployment debt compounds. Relative cost of remediating a seam rises steeply with the stage at which it is addressed (illustrative).}
\label{fig:debt}
\end{figure}

\textbf{Financial debt.} The cost of remediating a seam rises sharply the later it is addressed. A data-access boundary that is trivial to design correctly at the outset becomes expensive to retrofit once a system is in production, and a post-incident cost once it has failed in front of users or regulators (Fig.~\ref{fig:debt}). Deferring seam work does not avoid the cost; it multiplies it.

\textbf{Organizational debt.} When workflows are wrapped rather than redesigned, and when skills are not built, every subsequent initiative inherits the same unpaid friction. The organization does not learn to cross the seams; it re-encounters them, project after project, paying the interest each time.

\textbf{Competitive debt.} While an organization re-runs pilots against seams it has not resolved, rivals that have crossed those seams compound their advantage in cost, speed, and organizational learning. The gap widens not linearly but geometrically, because each cleared initiative makes the next one cheaper for the leader and no cheaper for the laggard. The true hazard of the current moment is therefore not a single failed pilot on this quarter's statement; it is a balance sheet quietly accumulating deployment debt while competitors retire theirs.

Crucially, deployment debt can be estimated, and the estimate is itself a management tool. For each seam, an organization records whether it is unresolved, partially resolved, or resolved, and attaches to each unresolved seam an order-of-magnitude remediation cost reflecting the stage at which it will likely be addressed. Summed across an initiative's open seams, the result is a deployment-debt figure that can be tracked over time and across the portfolio, much as technical debt is tracked in mature engineering organizations \citep{breck2017}. The precise numbers matter less than the discipline of making the liability visible; a debt that is named is a debt that can be prioritized and retired.

\section{The Seam Index: A Diagnostic Instrument}
\label{sec:index}
What gets measured gets managed. The Seam Index operationalizes the seam analysis into a single, decision-grade score. For each of the six seams, a platform is scored on a three-point scale: 0 if the adopter must build and govern the seam itself, 1 if the platform provides partial or assisted support, and 2 if the seam is inherited natively from the platform. Summed across the six seams, the Index yields a 0--12 score (Table~\ref{tab:rubric}).

\begin{table}[t]
\centering
\small
\caption{The Seam Index scoring rubric (0--2 per seam; 0--12 total).}
\label{tab:rubric}
\begin{tabularx}{\linewidth}{L{1.1cm} X L{3.4cm}}
\toprule
\textbf{Score} & \textbf{Meaning} & \textbf{Interpretation for the buyer} \\
\midrule
0 & The adopter must build, integrate, and govern the seam from scratch. & Full deployment-debt exposure; consulting-intensive. \\
1 & The platform provides partial or assisted capability for the seam. & Shared burden; some residual friction remains. \\
2 & The seam is inherited natively from the platform and its ecosystem. & Friction removed; the adopter starts past the seam. \\
\bottomrule
\end{tabularx}
\end{table}

\subsection{Scoring protocol and evidence anchors}
To make the instrument reproducible rather than impressionistic, each score must be justified against a documented artifact rather than a vendor claim. The protocol is: (1)~for each seam, the evaluator records the concrete evidence required for a score of~2 and for a score of~1; (2)~scoring is performed by a cross-functional group (data, security/identity, risk, and a process owner) rather than an individual, to reduce single-rater bias; (3)~each score cites an artifact---an architecture diagram, a signed control, a configuration, or a contract term---so the score is auditable. For example, the identity seam earns a 2 only on documented native single-sign-on and role-based-access integration with the organization's existing identity provider; a 1 on partial or connector-based support; and a 0 when access control must be built by the adopter. Analogous anchors are defined for the remaining seams (native data connectors and lineage for data; built-in audit logging and monitoring for governance; and so on). This mirrors the rubric philosophy of production-readiness scores in ML engineering \citep{breck2017}, applied to the platform-selection decision rather than to a single system.

\begin{figure}[t]
\centering
\includegraphics[width=0.6\linewidth]{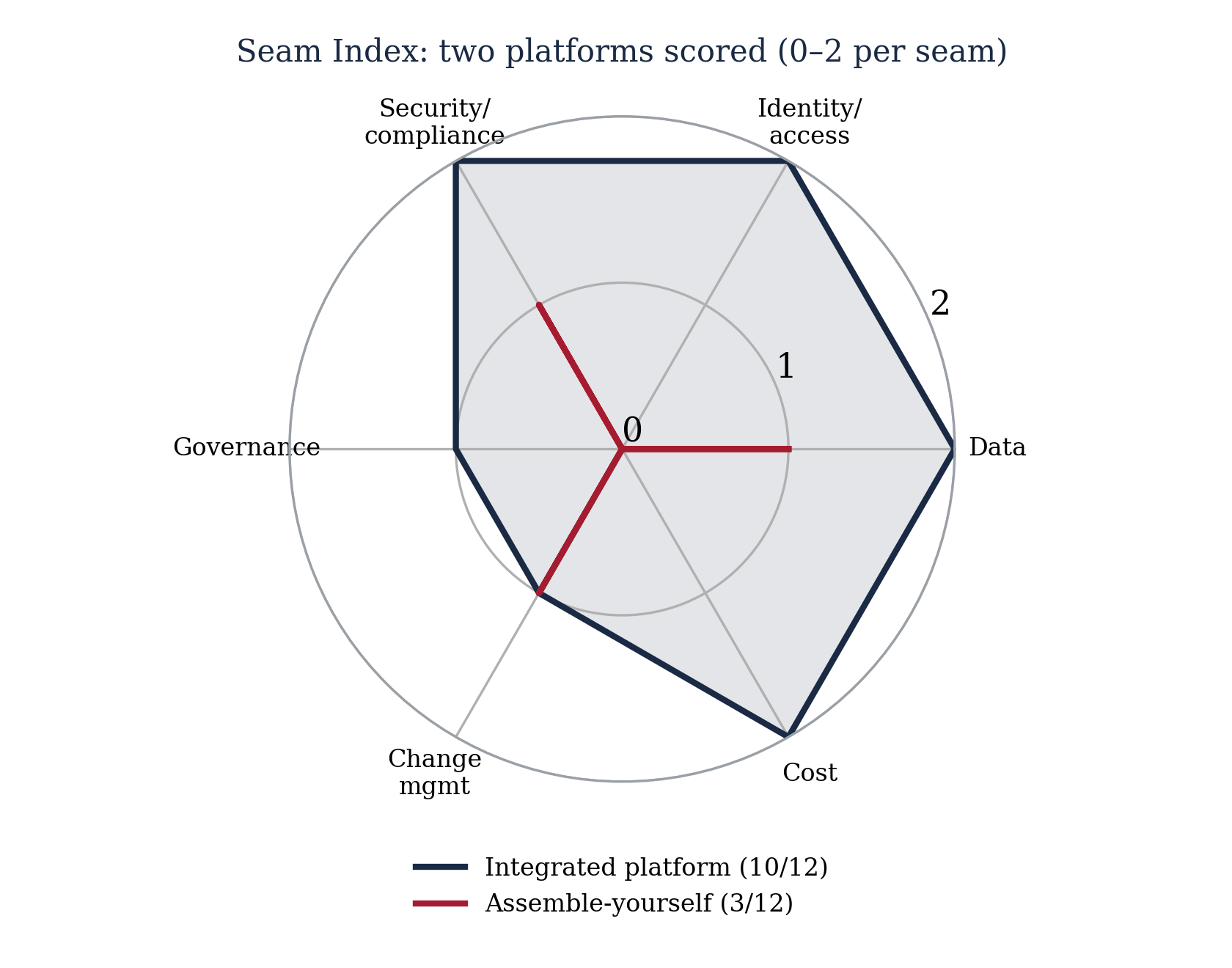}
\caption{The Seam Index applied. Two platforms scored across the six seams; the integrated platform removes far more friction (illustrative worked example).}
\label{fig:radar}
\end{figure}

The Index reframes the buying decision. Instead of asking which platform posts the highest benchmark---an attribute that is converging and swappable---the manager asks a question the benchmark cannot answer: how many seams does this platform remove from the environment we already operate? Figure~\ref{fig:radar} shows two platforms scored on the same six seams. An integrated platform that inherits identity, data, and governance from an existing enterprise estate may score 10 of 12, while a nominally more ``open'' platform that leaves those seams to the buyer scores 3 of 12---even if the second hosts a marginally stronger model.

The available evidence rewards the answer the Index points toward. Field research finds that purchased, integrated solutions reach production roughly twice as often as internally built ones \citep{nanda2025}, consistent with the interpretation that buying a platform is, in seam terms, buying the removal of friction the organization would otherwise finance itself (Table~\ref{tab:buybuild}).

\begin{table}[t]
\centering
\small
\caption{Buy versus build, in seam terms.}
\label{tab:buybuild}
\begin{tabularx}{\linewidth}{L{3.0cm} X L{3.3cm}}
\toprule
\textbf{Path} & \textbf{Seam burden} & \textbf{Observed outcome} \\
\midrule
Buy an integrated platform & Many seams inherited; lower deployment-debt exposure. & Reaches production roughly twice as often \citep{nanda2025}. \\
Build internally & Most seams owned by the adopter; higher deployment debt. & Lower production rate; higher long-run cost \citep{nanda2025,menlo2025}. \\
\bottomrule
\end{tabularx}
\end{table}

\subsection{Worked example: applying the Index end to end}
Consider a composite but representative decision: a regulated insurer choosing between two platforms for a claims-adjudication assistant. Platform~A hosts a model ranked second on public benchmarks but is delivered inside the productivity and identity suite the insurer already runs; Platform~B hosts the top-ranked model but must be integrated from scratch. Scored on the Index, Platform~A inherits identity~(2), data perimeter~(2), governance~(2), and security~(2), with partial change-management tooling~(1) and partial cost controls~(1), for a total of~10. Platform~B, despite its stronger model, scores identity~(0), data~(1), governance~(0), security~(1), change~(0), and cost~(0), for a total of~2. The Deployment Wall predicts that Platform~B's initiative will leak value precisely at integration, governance, and adoption---the seams it leaves to the insurer---while Platform~A begins several courses up the wall. The benchmark gap that dominates the vendor pitch is, in seam terms, the least decision-relevant fact on the table.

Two caveats keep the Index honest. First, scoring involves judgment, and two evaluators may differ on whether a seam is partially or fully inherited; the remedy is the evidence-anchored, group-scored protocol above, and reporting inter-rater agreement when the Index is used in research. Second, the Index measures friction removed, not fitness for a specific task; a platform can score well and still be wrong for a particular use case. The Index is thus a necessary filter, not a sufficient one: it eliminates options that will drown in deployment debt, leaving a shortlist on which task fit and cost can be judged.

\section{The New Enterprise AI Moat}
\label{sec:moat}
Follow the logic to its conclusion and the competitive picture inverts. If intelligence is converging and swappable, the model can no longer be the moat. The durable advantage is friction removed---and it accrues to whichever provider owns the most seams for a given enterprise environment. This is the complementary-assets logic of \citet{teece1986} applied to AI: when the core technology is imitable, returns flow to the co-specialized assets, which here are the seams. Crucially, that is not a single company. The moat is distributed, because different hyperscalers own different seams (Table~\ref{tab:moat}).

\begin{table}[t]
\centering
\small
\caption{The moat has moved to the seam layer, and it is distributed.}
\label{tab:moat}
\begin{tabularx}{\linewidth}{L{2.0cm} L{3.0cm} X}
\toprule
\textbf{Provider} & \textbf{Seam it owns} & \textbf{Source of advantage} \\
\midrule
Microsoft & Distribution \& identity & Hundreds of millions of existing productivity seats and enterprise identity. \\
Amazon & Data perimeter & Enterprise data already resident in its cloud, inside the security boundary. \\
Google & Vertical integration & Lab, model, governance plane, and silicon under one roof. \\
\bottomrule
\end{tabularx}
\end{table}

The distribution of the moat explains why no single provider ``wins'' enterprise AI outright. Microsoft's advantage is distribution and identity: existing productivity seats and a dominant enterprise identity fabric mean that, for many organizations, its AI capabilities arrive already past the identity and change-management seams. Amazon's advantage is the data perimeter: for enterprises whose data already resides in its cloud, the data and security seams are substantially pre-crossed. Google's advantage is vertical integration: owning the lab, the model, the governance plane, and the silicon lets it close seams in a coordinated way that assemblers of third-party components cannot easily match. None of these advantages is the model itself; each is a seam the provider has already removed for a particular class of customer---consistent with the AI-factory view of competition \citep{iansiti2020}.

\begin{figure}[t]
\centering
\includegraphics[width=0.66\linewidth]{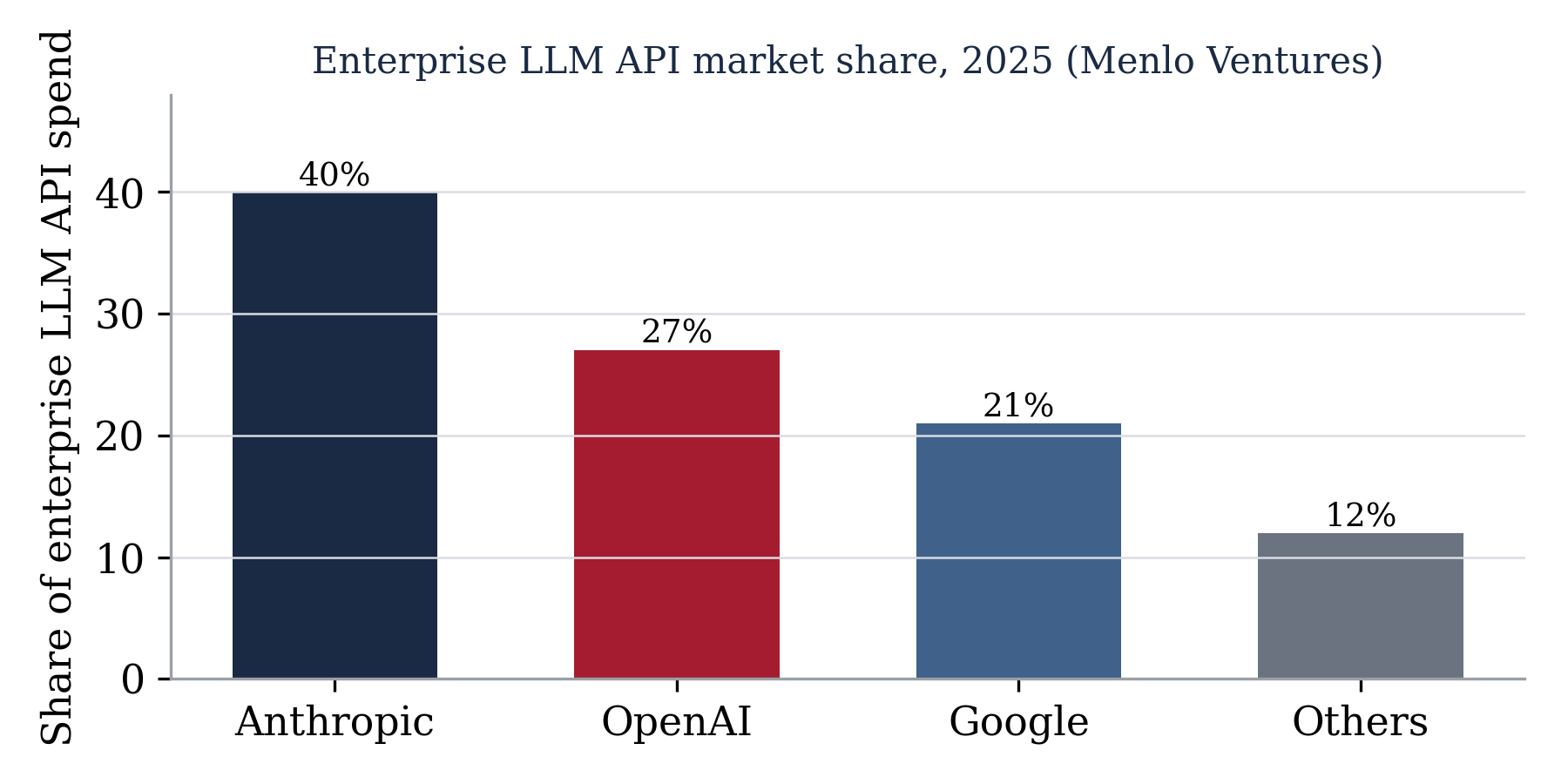}
\caption{Enterprise LLM API market share is split across providers, consistent with a multi-model norm (2025 data).}
\label{fig:share}
\end{figure}

A further consequence is the rising strategic value of the \emph{orchestration layer}---the control plane through which an enterprise routes requests across several models while enforcing a single set of identity, data, and governance rules. In a multi-model world, whoever owns that layer owns the seams that matter, because the models beneath it are interchangeable by design. The multi-model reality reinforces this: recent spend data indicate that roughly 79\% of Anthropic's enterprise customers also pay for OpenAI \citep{ramp2026}, and the enterprise application-programming-interface (API) market is now split across several providers rather than dominated by one (Fig.~\ref{fig:share}). If the specific model is interchangeable, advantage belongs to the platform that lets an enterprise orchestrate many external models while inheriting identity, data perimeter, and governance from a single, trusted control plane. The model layer commoditizes; the seam layer is the moat.

\section{Implications for Practice}
\label{sec:practice}
The framework converts directly into managerial action. Table~\ref{tab:action} lists five moves a leader can make immediately, each tied to a construct developed above.

\begin{table}[t]
\centering
\small
\caption{From framework to action.}
\label{tab:action}
\begin{tabularx}{\linewidth}{L{3.0cm} X L{3.0cm}}
\toprule
\textbf{Action} & \textbf{What it means} & \textbf{Rationale} \\
\midrule
Score the seams first & Apply the Seam Index to every option before the next vendor demo. & Reframes the decision from benchmark to architecture (Sec.~\ref{sec:index}). \\
Run a deployment-debt audit & Quantify unresolved friction on stalled initiatives as financial, organizational, and competitive debt. & Makes hidden liability visible to finance (Sec.~\ref{sec:debt}). \\
Redesign work, don't wrap it & Pair architects with the operators who own each process; change the workflow. & Change management is the seam most correlated with value (Sec.~\ref{sec:seams}). \\
Buy over build for non-differentiators & Inherit seams for anything that is not a source of competitive advantage. & Purchased platforms reach production $\sim$2$\times$ as often (Sec.~\ref{sec:index}). \\
Treat consulting intensity as a signal & Read heavy deployment-services requirements as high seam burden. & The labs' own behavior reveals where the constraint lies (Sec.~\ref{sec:tell}). \\
\bottomrule
\end{tabularx}
\end{table}

The urgency of these moves is compounded by the approaching agentic wave. Agentic systems, which chain model calls into autonomous multi-step actions, do not reduce seam burden---they amplify it, because autonomous action across identity, data, and governance boundaries raises the stakes at every seam. Industry analysts already forecast that a large share of agentic-AI initiatives will be canceled before reaching production \citep{gartner2025}. An organization that has not learned to cross the six seams for a single-step assistant will not cross them for an autonomous agent; it will simply fail at greater scale and cost. The Deployment Era's disciplines are therefore not optional preparation for the agentic future---they are its prerequisite.

\subsection{Organizing for deployment}
An organization that internalizes the Deployment Era reorganizes around seams rather than around models. In practice this means standing up a small, cross-functional deployment function that owns the six seams as first-class concerns: data engineers and stewards for the data seam; identity and security specialists for the access and compliance seams; a governance owner accountable to risk and audit; and change leaders paired with process owners for adoption. The model itself becomes a swappable input to this function rather than its centerpiece. The organizations that field research identifies as capturing value are disproportionately those that have built this durable capability rather than those that have simply run the most pilots \citep{nanda2025,mckinsey2025}---an instance of the firm-level AI capability the IS literature has begun to formalize \citep{mikalef2021}.

\subsection{Sequencing the agentic transition}
Because agentic architectures act autonomously across identity, data, and governance boundaries, they raise the stakes at every seam rather than lowering them; an agent that can take actions must be governed more tightly than an assistant that only suggests them. The prudent sequence is to close the six seams for assisted use first, prove the governance and identity controls under lower stakes, and only then extend autonomy. Organizations that invert this order---granting autonomy before the seams are crossed---are the ones most exposed to the wave of cancellations analysts anticipate \citep{gartner2025}.

\section{Propositions and a Research Agenda}
\label{sec:props}
To move the framework from argument to testable theory, we offer six propositions. Each is stated in falsifiable form and is amenable to empirical study using platform-level and initiative-level data.

\begin{table}[t]
\centering
\small
\caption{Testable propositions.}
\label{tab:props}
\begin{tabularx}{\linewidth}{L{0.8cm} X}
\toprule
\textbf{\#} & \textbf{Proposition} \\
\midrule
P1 & The probability that an enterprise AI initiative reaches production is more strongly associated with the platform's Seam Index than with the underlying model's benchmark rank. \\
P2 & Higher Seam Index scores are associated with lower total cost of ownership over the initiative lifecycle. \\
P3 & The cost of remediating a given seam increases monotonically with the deployment stage at which it is first addressed (deployment debt compounds). \\
P4 & Organizations that redesign workflows (vs.\ wrapping them) realize measurable value at a higher rate, controlling for model quality. \\
P5 & Reliance on external deployment consulting is positively associated with platform seam burden (low Seam Index). \\
P6 & In multi-model enterprises, competitive advantage concentrates in providers that own identity, data-perimeter, and governance seams rather than in providers of the highest-ranked model. \\
\bottomrule
\end{tabularx}
\end{table}

A productive research agenda would construct a validated, multi-item Seam Index instrument and test its inter-rater reliability; assemble a cross-industry panel of initiatives to estimate the associations in P1 and P4; and study deployment debt longitudinally to calibrate the compounding function in P3. Because the constructs are defined operationally, each proposition can be examined with a combination of survey, archival, and case-based methods. Beyond the individual propositions, a broader program could test whether the Seam Index behaves as a \emph{leading indicator}: whether a platform's score at the point of selection predicts production success months later better than any model-quality metric available at the same moment. Establishing that predictive relationship---or falsifying it---would settle the central empirical question this paper raises: whether, in the current era, friction removed predicts enterprise AI value better than intelligence acquired.

\section{Limitations}
\label{sec:limits}
Three limitations bound the contribution. First, the framework is conceptual and practitioner-derived; its propositions are offered for testing, not as validated findings, and the field observations that motivated it are non-random and drawn from the author's own engagements. Second, several quantitative figures are illustrative models that encode the direction and approximate magnitude of cited evidence rather than measured datasets; readers should treat them as communicative devices. Third, the empirical landscape is moving quickly: market-share, spend, and partnership figures are current to the time of writing and should be re-verified before they are relied upon, as the underlying arrangements among labs, hyperscalers, and integrators are evolving month to month.

\section{Conclusion}
\label{sec:concl}
The Intelligence Era rewarded whichever organization could reach the most capable model---an advantage that necessarily erodes once every organization has access to a model that is good enough. The Deployment Era rewards something far harder to copy: the organizational and architectural capability to move any capable model through the six seams and into the work. This paper has argued that the wall enterprises keep hitting is built of deployment friction, not model quality; that the friction can be located (the Deployment Wall), measured (the Seam Index), and costed (deployment debt); and that the durable moat has migrated to whichever provider removes the most friction for a given environment. The managerial task is correspondingly clear. Leaders should stop scoring AI on benchmarks alone and start scoring platforms on the seams they remove, redesign the work rather than wrap it, and retire deployment debt before it compounds. The winners of the next decade will not be the organizations that waited for a smarter model; they will be the ones that treated deployment as the discipline it has become.

\section*{Acknowledgments}
The author thanks the enterprise, mid-market, and startup teams whose deployment engagements informed the observations in this paper; specific organizations are not named to preserve confidentiality. This research received no external funding. The author used generative-AI tools to assist with literature synthesis and with drafting and editing; all ideas, analysis, and conclusions are the author's own, and the author is solely responsible for the content.

\bibliographystyle{plainnat}
\bibliography{refs}

\end{document}